\documentclass[aps, prc, 10pt, twocolumn, showpacs]{revtex4}
\usepackage{hyperref}
\usepackage{natbib}
\usepackage{graphicx}
\usepackage{txfonts}
\usepackage[dvipsnames,usenames]{xcolor}
\usepackage{soul}
\usepackage[normalem]{ulem}

\newcommand*{\ndrip}{\ensuremath{n_{\mathrm{drip}}}}

\newcommand*{\dif}{\ensuremath{\mathrm{d}}}
\newcommand*{\gauss}{\ensuremath{\,\mathrm{G}}}
\newcommand*{\Msun}{\ensuremath{\,M_{\odot}}}
\newcommand*{\fermi}{\ensuremath{\,\mathrm{fm}}}
\newcommand*{\invcfm}{\ensuremath{\fermi^{-3}}}
\newcommand*{\km}{\ensuremath{\,\mathrm{km}}}
\newcommand*{\Hz}{\ensuremath{\,\mathrm{Hz}}}
\newcommand*{\MeV}{\ensuremath{\,\mathrm{MeV}}}
\newcommand*{\vAlfven}{\ensuremath{\varv_{\mathrm{A}}}}
\newcommand*{\vshear}{\ensuremath{\varv_{\mathrm{s}}}}
\newcommand*{\fent}{\ensuremath{f_{\mathrm{ent}}}}
\newcommand*{\wCoul}{\ensuremath{w_{\mathrm{Coul.}}}}
\newcommand*{\Eshell}{\ensuremath{E_{\mathrm{shell}}}}
\newcommand*{\Esurf}{\ensuremath{E_{\mathrm{surf}}}}
\newcommand*{\Epair}{\ensuremath{E_{\mathrm{pair}}}}
\newcommand*{\Ebind}{\ensuremath{E_{\mathrm{bind}}}}

\newcommand*{\apjfigscale}{0.85}
\newlength{\apjcolwidth}
\newlength{\figwidth}
\newlength{\doublewide}
\begin{document}
 
\title{Magnetar giant flare oscillations and the nuclear symmetry energy}
 
\author{Alex T. Deibel\,}

 \email[]{deibelal@msu.edu}
\affiliation{Department of Physics and Astronomy, Michigan State
  University, East Lansing, MI 48824, USA}
\affiliation{The Joint Institute for Nuclear Astrophysics}
\author{Andrew W. Steiner}
\affiliation{The Joint Institute for Nuclear Astrophysics}
\affiliation{Institute for Nuclear Theory, University of Washington, Seattle, WA 98195, USA}
\author{Edward F. Brown}
\affiliation{Department of Physics and Astronomy, Michigan State University, East Lansing, MI 48824, USA,}
\affiliation{The Joint Institute for Nuclear Astrophysics}
\affiliation{National Superconducting Cyclotron Laboratory, Michigan State University, East Lansing, MI 48824, USA}

\begin{abstract}
If the observed quasi-periodic oscillations in magnetar flares are partially
  confined to the crust, then the oscillation frequencies are unique
  probes of the nuclear physics of the neutron star crust. We study
crustal oscillations in magnetars including corrections for a finite
Alfv\'en velocity. Our crust model uses a new nuclear mass formula
that predicts nuclear masses with an accuracy very close to that of
the finite range droplet model. This mass model for equilibrium nuclei
also includes shell corrections and an updated neutron-drip line. We
perturb our crust model to predict axial crust modes and assign them
to observed giant flare quasi-periodic oscillation frequencies from
the soft $\gamma$-ray repeater
SGR 1806-20. We find magnetar crusts that match observations for
various magnetic field strengths, entrainment of the free neutron gas
in the inner crust, and crust-core transition densities. We find
  that observations can be reconciled with smaller values of the
  symmetry energy slope parameter, $L$, if there is a significant
  amount of entrainment of the neutrons by the superfluid or if the
  crust-core transition density is large.
  We also find neutron star masses and radii which are in
    agreement with expectations from what is known about low-density
    matter from nuclear experiment. Matching observations with a
  field-free model we obtain the approximate values of $M =1.35\Msun$
  and $R = 11.9\km$. Matching observations using a model with the
  surface dipole field of SGR 1806-20 ($B=2.4\times10^{15}\gauss$) we
  obtain the approximate values of $M = 1.25\Msun$ and $R = 12.4\km$.
 \end{abstract}

\preprint{INT-PUB-13-011}
 
 \maketitle
 
\section{Introduction}

Highly magnetized and isolated neutron stars, known as magnetars, emit
irregular and extremely energetic $\gamma$-ray flares. These flares are
thought to occur following a starquake, in which a reconfiguration of
the magnetic field fractures the magnetar's crust. Quasi-periodic
oscillations (QPOs) are observed in the tails of giant flare emissions
\citep{barat83, israel05, strohmayer05, watts06}. Following the
proposal by \citet{duncan98}, many have attempted to model the QPOs as
torsional modes of the crust \citep[e.g.,][]{piro05, strohmayer06,
  samuelsson07}. If this is indeed the cause of the QPOs, then
magnetars can give a unique insight into the microphysics of the
neutron star crust \citep{steiner09}, e.g., the nuclear symmetry
energy $S(n)$, here defined as the difference in energy between pure
neutron matter and proton-neutron symmetric matter as a function
  of the baryon density, $n$. In particular, crust frequencies are
  sensitive to the quantity $L\equiv 3 n_0
  \left(\partial S/\partial n\right)_{n=n_0}$~\citep{steiner09,sotani13}. In addition, different
modes have different scalings with the neutron star mass and radius;
it follows that observations of two or more modes, such as a
fundamental and harmonic, can constrain the magnetar's mass and radius
\citep{samuelsson07,lattimer07} and hence the equation of state (EOS)
of dense matter.

In this paper, we explore the impact on constraints of the symmetry energy and
  the properties of the crust under the assumption that one of the
lower frequency QPOs and a higher frequency QPO, such as the 626~Hz
mode observed from the soft $\gamma$-ray repeater SGR 1806-20 \citep{watts06}, represent the
fundamental and first harmonic modes of the crust. Our study extends
previous work by using a modern EOS to predict torsional mode
frequencies and by matching observations for various magnetic field
strengths, entrainment fractions for free neutrons in the inner crust,
and crust-core transition densities. The crust EOS is based on a
liquid droplet model which predicts nuclear masses to within $1.2\MeV$
\citep{steiner12b}, close to the accuracy obtained in the finite range
droplet model \citep{moller95}. By adding the effect of the magnetic
field on electrons, as in \citet{broderick00}, we revise the magnetic
composition of the crust \citep{lai91} with a new determination of
equilibrium nuclei. The core EOS is based on recent neutron star mass
and radius constraints from observations of photosphere radius
expansion bursts (PREs) and the quiescent emission of low-mass x-ray
binaries (LMXBs) \citep{steiner-lattimer-brown-10,steiner12}.

While this initial study incorporates a modern EOS, it
necessarily makes many simplifying assumptions. The basis of our
approach is the assumption that the QPOs in question are indeed due to
torsional crust modes. An alternative model associates the QPOs with
magnetohydrodynamic (MHD) modes in the core
\citep{glampedakis06,levin06}. We note that neither model is able to
predict all of the observed mode frequencies. Crustal oscillations
cannot easily reproduce all of the low-frequency modes
\citep{samuelsson07}. Recent studies of core MHD modes with crust-core
coupling in a dipolar magnetic field found that core MHD modes could
explain most of the QPOs, but only with a magnetic field larger than the
observed surface dipole fields \citep{gabler12}. In addition, core MHD
models have been unable to reproduce the highest QPO frequencies
observed \citep{vanHoven12}. Stratification and entrainment can
increase core MHD frequencies, but this does not yet completely
explain the data either \citep{Passamonti12}. We also do not
include a consistent treatment of the nuclear pasta or its shear viscosity. We comment on how this might modify our results at the end.
 
With these caveats, we construct magnetized crust models for
equations of state with extreme values of $L$,
free neutron entrainment, crust-core transition densities, and
  different high-density equations of state. In Sec.~\ref{s.crust-composition} we
present the magnetized crust composition based on our mass model (a
detailed description of this formalism is in
Appendix~\ref{s.crustEOS}). Section~\ref{s.torsional-oscillations}
contains a summary of the axial perturbation equations for the crust
modes. We then use, in Sec.~\ref{s.results}, the predicted
fundamental and harmonic frequencies, along with the magnetized crust
composition, to investigate the role of the nuclear symmetry energy in determining
magnetar masses, radii, and crust oscillation frequencies. In Sec.~\ref{s.discussion}, 
we discuss our results.

\section{Crust Composition} \label{s.crust-composition}

For an isolated neutron star we determine the crust composition by
finding the ground-state nucleus at a given baryon density $n$. The
outer crust consists of a lattice of nuclei embedded in a degenerate
electron Fermi gas \citep{baym71}. The neutron-drip point, the point
at which it becomes energetically favorable for neutrons to exist
outside of the nuclei, defines the boundary between the outer and
inner crust. The inner crust can then be described as a lattice of
nuclei embedded in both an electron and neutron gas \citep{baym71}.
We use a liquid droplet model with ``squared-off'' nuclear
  density distributions~\citep{1985NuPhA.432..646L} and assume that
  the number density of neutrons and protons inside nuclei is fixed
  and that the number density of neutrons outside nuclei is fixed in a
  way as to obtain equilibrium. The nuclei occupy a fraction, $\chi$,
  of the total volume of a Wigner-Seitz cell and the dripped neutrons
  in the inner crust occupy a fraction, $1-\chi$, of the total
  volume. This separation of the neutron gas from the nuclei is
  convenient for modeling the properties of matter at high densities
  \citep{steiner12b}. As described in
Appendix~\ref{s.crustEOS}, at a given $n$, proton number $Z$, and
atomic number $A$, the total energy density of the crust will have
contributions from the nuclear binding energy, the Coulomb lattice,
the electron gas, and the neutron gas. Both the bulk energy of the
nucleus and the energy of the neutron gas are determined by the same
Skyrme interaction, either SLy4 with $L=46\MeV$ or Rs with $L=86\MeV$.
The liquid drop model parameters \citep{steiner12b} are fit to experimental nuclear
masses separately for each interaction. The fits for
  each interaction differ because the value of $L$ determines the
  surface energy and surface symmetry energy parameters. Both
  models give similar quality fits to the data. \citet{chamel12b} obtained qualitatively similar results using
Hartree-Fock-Bogoliubov models.

\begin{figure}
\centering
\includegraphics[width=\figwidth]{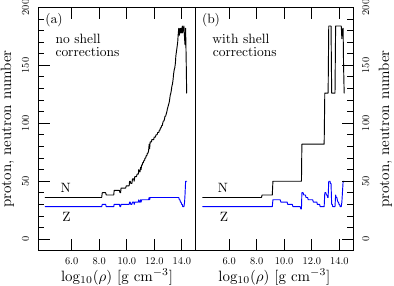}
\caption{(Color online) Equilibrium composition of the crust without a magnetic field
  for (a) a model that neglects shell effects and (b) a model
  that includes them. The blue lower curve corresponds
  to the proton number $Z$ and the black upper curve corresponds to
  the neutron number $N$. 
\label{f.composition}}
\end{figure}

At a given $n$ the equilibrium nucleus minimizes the total
energy density of the system. The most energetically favorable nuclei
tend to contain a closed shell of protons or neutrons due to shell
corrections~\citep{sato79, dieperink09}. As $n$ increases,
equilibrium nuclei will move to higher closed shells of protons and
neutrons with the most neutron-rich nuclei seen between the
neutron-drip point and the crust-core transition. The above features
of the crust composition can be seen in Fig.~\ref{f.composition}. We
ignore the deformation of nuclei at high densities in the crust
composition.
 
  In the outer crust a strong magnetic field will force electrons to
occupy the lowest Landau levels. At higher baryon densities electrons
can occupy higher Landau levels and thus their energy density
approaches the field-free case. For this reason, as seen in
Table~\ref{t.composition}, only the outer crust equilibrium
composition is significantly altered. For $B<10^{18}\gauss$ we can
ignore both the effect of the magnetic field on the structure of the
nuclei in the crust \citep[see, e.g.,][]{harding06,nag09} and on the
gross structure of the neutron star~\citep{cardall01}.

\begin{table*}
\caption{Magnetic Equilibrium Nuclei Below Crust-Core Transition\label{t.composition}}
\centering
\renewcommand{\arraystretch}{1.4}
\begin{ruledtabular}
\begin{tabular}{cccccc}
 & \multicolumn{5}{c}{$\rho_{\max}$ ($\mathrm{g\,cm}^{-3}$)}\\
\cline{2-6}
Nuclei\footnotemark[1] & $B_* = 0$ & $B_* = 1$ & $B_* = 10$ & $B_* = 10^2$ & $B_* = 10^3$ \\
\hline
$^{64}_{28}\mathrm{Ni}\ $ & $2.23\times10^{8}$ &  & $2.33\times10^{8}$  & $1.63\times10^{9}$  & $1.75\times10^{10}$ \\
$^{66}_{28}\mathrm{Ni}\ $ & $1.37\times10^{9}$ &  & $1.40\times10^{9}$  & $2.92\times10^{9}$ & $2.71\times10^{10}$ \\
$^{84}_{34}\mathrm{Se}\ $ & $5.66\times10^{9}$ & & & $4.87\times10^{9}$ & $5.29\times10^{10}$  \\
$^{82}_{32}\mathrm{Ge}\ $ & $1.73\times10^{10}$ & &  & $1.69\times10^{10}$ & $7.62\times10^{10}$ \\
$^{80}_{30}\mathrm{Zn}\ $ & $3.99\times10^{10}$ & & &$3.94\times10^{10}$ & $1.01\times10^{11}$  \\
$^{78}_{28}\mathrm{Ni}\ $ & $1.56\times10^{11}$ & &  & $1.57\times10^{11}$  & $1.61\times10^{11}$  \\
$^{76}_{26}\mathrm{Fe}\ $ & $1.86\times10^{11}$ & &  & $1.85\times10^{11}$ & $1.76\times10^{11}$  \\
$^{122}_{40}\mathrm{Zr}\ $ & $2.51\times10^{11}$ & & & $2.52\times10^{11}$ & $1.98\times10^{11}$  \\
$^{120}_{38}\mathrm{Sr}\ $ & $3.54\times10^{11}$ & &  &$3.54\times10^{11}$ & $4.04\times10^{11}$  \\
$^{118}_{36}\mathrm{Kr}\ $ & $5.17\times10^{11}$ & & &$5.15\times10^{11}$ & $5.77\times10^{11}$  \\
$^{116}_{34}\mathrm{Se}\ $ & $8.11\times10^{11}$ & & & $8.13\times10^{11}$ & $8.56\times10^{11}$ \\
$^{114}_{32}\mathrm{Ge}\ $ & $2.35\times10^{12}$ & & &  & $2.25\times10^{12}$ \\
$^{112}_{30}\mathrm{Zn}\ $ & $3.94\times10^{12}$ & & & & $4.02\times10^{12}$ \\
$^{110}_{28}\mathrm{Ni}\ $ & $8.64\times10^{12}$ & & & & $8.65\times10^{12}$ \\
$^{166}_{40}\mathrm{Zr}\ $ & $1.07\times10^{13}$ & & & & $1.08\times10^{13}$\\
\end{tabular}
\end{ruledtabular}
\footnotetext[1]{We adopt the format of \citet{lai91} where
  $\rho_{\max}$ is the maximum mass density where the equilibrium
  nucleus is present. If a density value is unchanged the following
  column is blank. Here $B_* = B/(4.414 \times 10^{13}\gauss$), which
  is the ratio of the magnetic field to the critical field, defined as
  the field at which the cyclotron energy equals the electron
  rest-mass.}
 \end{table*}

\section{Torsional Oscillations in a Strong Magnetic Field}
\label{s.torsional-oscillations}

\begin{figure}[htbp]
\includegraphics[width=\figwidth]{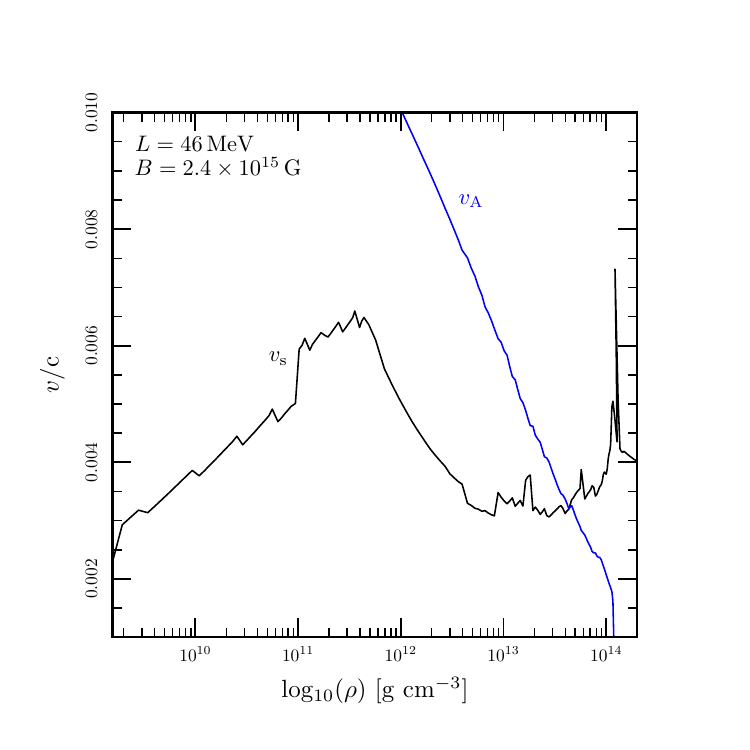}
\caption{(Color online) Alfv\'en velocity (blue curve) 
  and shear velocity (black curve)
  in the crust as a function of mass density. The composition
  is that of a $1.4\Msun$ neutron star using the SLy4 crust EOS.
}
\label{f.Alfven-shear-speeds}
\end{figure}

We describe the axial crust modes of an oscillating neutron star in
the relativistic Cowling approximation following the work of
\citet{schumaker83} and \citet{samuelsson07}. We combine two forms of
the axial perturbation equation. In the non-magnetic case, the
equation for the axial perturbation $\xi$ can be written in the form
\citep{samuelsson07} $\xi^{\prime\prime} + F^{\prime}\xi^{\prime} +
G\xi = 0$, in which primes indicate derivatives with respect to the
radial coordinate, and $F$ and $G$ are functions of the shear velocity
$\vshear$ and the metric functions $\nu$ and $\lambda$ for a static and
spherically symmetric space-time metric, $\dif s^{2} = -e^{2\nu}\dif
t^{2} + e^{2\lambda}\dif r^{2} + r^{2} \left(\dif \theta^{2} +
\sin^{2}\theta\,\dif\phi^{2}\right)$. Because of the strong vertical
stratification, radial perturbations are driven to small amplitudes
and high frequencies and are much less relevant in the crust. Working
in the isotropic limit, we incorporate corrections for a finite
Alfv\'en velocity $\vAlfven = B/\sqrt{4\pi\rho_{i}}$ in the radial
direction by analogy with the Newtonian expressions \citep{piro05,
  steiner09}. A more complicated magnetic field configuration is
worthy of study, but we find below that crustal frequencies are more
sensitive to entrainment and the EOS and thus we choose a uniform
magnetic field for now.  The result is
\begin{eqnarray}
\lefteqn{(\vshear^{2} + \vAlfven^{2})\xi^{\prime\prime} 
+ \vshear^{2}\frac{\dif}{\dif r} \left\{ \ln \left[r^4 e^{\nu-\lambda} 
\left(\varepsilon + p\right) \varv_s^2\right] \right\} 
\xi^{\prime} } \nonumber\\
&&{} + e^{2\lambda}\left[e^{-2\nu}\omega^{2} 
\left(1 + \frac{\vAlfven^{2}}{c^{2}}\right) -	
\frac{\left(l^{2}+l-2\right)\varv_s^{2}}{r^{2}}\right]\xi =  0 .
\label{e.perturb}
\end{eqnarray}
In this expression $r$ is the radius, $\varepsilon$ is the energy
density, $p$ is the pressure, $\omega$ is the angular frequency, and
$l$ is the angular wave number. The shear velocity is $\vshear =
\sqrt{\mu/\rho}$, which is plotted with the Alfv\'en velocity in
Fig.~\ref{f.Alfven-shear-speeds}. Here $\mu$ is the shear modulus,
for which we use the formulation appropriate for a body-centered cubic
lattice \citep{strohmayer91},
\begin{equation}\label{e.shear-modulus}
\mu = \frac{0.1194\Gamma}{1 +  0.595(\Gamma_0/\Gamma)^{2}}n_i k_{\mathrm{B}} T.
\end{equation}
We integrate Equation~(\ref{e.perturb}) over the solid crust, which
lies between the crust-core interface at $r = R_{\mathrm{core}}$ and
where the lattice melts at $r = R_{\mathrm{crust}}$. The melting
transition is determined by where the plasma coupling parameter
$\Gamma = (Ze)^{2}/ak_{\mathrm{B}}T = \Gamma_{\!\mathrm{melt}} = 175$
\citep{farouki93, Horowitz2010}. Here $a =
\left(3/{4{\pi}n_i}\right)^{1/3}$ is the radius of the Wigner-Seitz
cell, $Z$ is the atomic charge number, $n_{i}$ is the ion number
density, and the temperature is $T= 3.0 \times 10^8\,\mathrm{K}$.

For the boundary conditions needed to solve
Equation~(\ref{e.perturb}), we require the traction, $\xi^{\prime}$,
to vanish at the top and bottom of the crust. This is a good
approximation near the surface where pressure vanishes. The
description of matter near the crust-core transition is complicated by
the appearance of nuclear pasta. Since the quasi-free neutrons are
superfluid, assuming the traction vanishes at the crust-core boundary
may also be a good approximation. An additional impact of the
superfluid is that some fraction, $\fent$, of the quasi-free neutrons
are entrained with the nuclei \citep{chamel12,chamel12c}. We assume
zero traction at the crust-core transition and leave a more complete
description of matter at the highest densities to future work.

For a given $l$, Equation~(\ref{e.perturb}), when integrated over the
crust with the boundary conditions described here, has an eigenvalue
$\omega$ that is uniquely determined by the crust thickness $\Delta =
R_{\mathrm{crust}}-R_{\mathrm{core}}$ and the neutron star radius.
These in turn depend on the equation of state.

\section{The Nuclear Physics of the Crust}
\label{s.results}

QPO frequencies have been detected in two magnetars, SGR 1806$-$20 and
SGR 1900$+$14 \citep{israel05, strohmayer05, watts06, strohmayer06}.
The $29\Hz$ mode in SGR 1806$-$20 and the $28\Hz$ mode in SGR 1900$+$14
are often assumed to be the fundamental torsional modes, but an 18 Hz
mode was also observed in SGR 1806$-$20 and a lower frequency mode is
not ruled out by the 1900$+$14 data. SGR 1806$-$20 also showed a very
clear $626\Hz$ mode, possibly matching the first radial harmonic
($n=1$). Several other modes are observed between 50 and $200\Hz$, and
these can be matched with higher angular momentum harmonics, $l>1$.
However, the frequency spacing is small between the $l>1$ harmonics
and this makes matching observed modes to particular angular momentum
harmonics ambiguous. We exclude an analysis of the higher angular
momentum harmonics because they do not lead to superior mass and
radius constraints. 

\subsection{The equation of state}

For the core, we use the probability distribution for the EOS
determined by \citet{steiner12} from observations of PREs and from the
quiescent emission of LMXBs. We construct five
equations of state corresponding to the most probable mass-radius relation 
along with its 1- and 2-$\sigma$ lower and upper bounds.
Our core model is distinct from the interaction used to describe
  matter in the crust (Skyrme models SLy4 or Rs) because we wish to
  avoid the additional assumption that the physics of matter at low
  and high densities is correlated. If we were to use the SLy4 EOS in the core, the radius of
  a $1.4\Msun$ neutron star is $11.65\km$ \citep{Stone03}, a bit
  larger than the lower 1-$\sigma$ results from \citet{steiner12}. For the 
  Rs EOS, the radius of a $1.4\Msun$ neutron star is $13.04\km$, which is very similar to the largest radius implied by the mass and
  radius observations from \citet{steiner12}.
Each equation of state gives mass-radius combinations with different
crust thicknesses, and hence a unique fundamental mode ($n=0$) and
harmonic mode ($n=1$). We can constrain the masses and radii of
magnetars by matching predicted fundamental modes and harmonic modes
to observed QPOs. While general relativistic corrections tend to decrease the
frequencies, softer core equations of state with smaller radii tend to
increase the frequencies. Because of this latter effect, we get
frequencies which are larger than that obtained by \citet{steiner09}.
The $n=0$, $l=2$ mode corresponds to the $29\Hz$ QPO of SGR 1806-20.
Our model predicts $n=1$ harmonic modes near $600\Hz$ and we compare
these predicted modes with the $626\Hz$ QPO of SGR 1806-20.

\begin{figure}
\centering
\includegraphics[width=\figwidth]{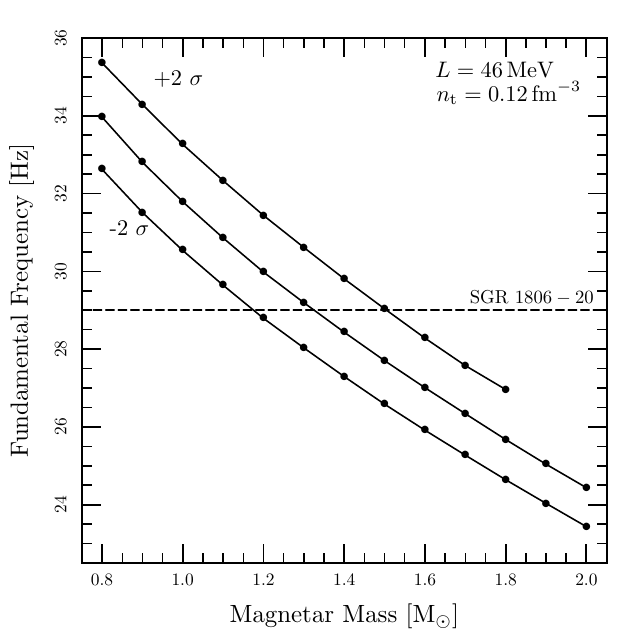}
\caption{Frequency of the fundamental $l=2$ mode as a function of the
  magnetar mass for the core EOS probability distribution (centroid
  and $\pm2\sigma$) from \citet{steiner12} and an SLy4 crust EOS. The
  dashed black line indicates the observed $29\Hz$ QPO of SGR 1806-20.
  The frequencies are evaluated for a crust-core transition density of
  $0.12\invcfm$ with $B = 0\gauss$.}
\label{f.frequency}
\end{figure}

To find crusts with fundamentals that match the $29\Hz$ QPO we model
crust perturbations in magnetars between $0.8\textrm{--}2.0\Msun$ with
magnetic fields matching the surface dipole field of SGR 1806-20.
Whichever crust has an $n=0$, $l=2$ mode that matches the $29\Hz$ QPO
we take as the crust of the magnetar. This method is demonstrated in
Fig.~\ref{f.frequency} where crusts are constructed using the SLy4
crust EOS~\citep{cha95}. Crusts with harmonics that match the $626\Hz$
QPO are found using an identical technique. We take whichever crust has an
$n=1$ mode that reproduces the observed QPO as the crust of the
magnetar. The same analysis is repeated for the 1 and 2-$\sigma$ lower
and upper bounds on the core EOS.

A comparison of masses and radii from fundamental and harmonic modes
can be seen in Fig.~\ref{f.mass-radius}. The intersection of
fundamental and harmonic masses and radii on the mass versus radius
plot gives a crust that best matches the properties of SGR 1806-20.
The mass and radius found for SGR 1806-20 depend on the properties of
the interior of the magnetar which determine the fundamental and
harmonic modes. This study focuses on varying three aspects of the
interior physics that remain unknown, namely, the magnetic field
strength in the crust, the crust-core transition density, and the
degree of free neutron entrainment in the inner crust.

 Different equations of state have different values of $L$ and thus different fundamental mode frequencies. 
The SLy4 EOS has a $29\Hz$ $n=0$, $l=2$ fundamental mode. 
The Rs EOS gives frequencies
between $15\textrm{--}20\Hz$ for the $n=0$, $l=2$ fundamental mode.
Therefore, we must assign the predicted fundamental modes from Rs to
the observed $18\Hz$ QPO. The Rs model has a smaller
fundamental frequency than SLy4 because its symmetry energy increases
more rapidly with density than does that of SLy4. With the density
dependence of the nuclear symmetry energy defined as $L\equiv 3 n_0
  \left(\partial S/\partial n\right)_{n=n_0}$, the Rs model has $L=86\MeV$ at
$n_0=0.16\invcfm$, the nuclear saturation density, whereas the SLy4 model has
$L=46\MeV$. The shear modulus is proportional to the plasma
coupling parameter $\Gamma$, which goes as $Z^2/a$ (see
Equation~(\ref{e.shear-modulus})). A larger value of $L$ tends to
increase the nucleon number $A$, which leads to larger $a$; the charge
number $Z$ is almost unchanged, however, due to nuclear shell effects.
Thus a larger value of $L$ decreases the shear modulus and also the
fundamental QPO frequency~\citep{steiner09, sotani13}.

\subsection{The crust magnetic field}

\begin{figure*}
\centering
\includegraphics[width=0.48\doublewide]{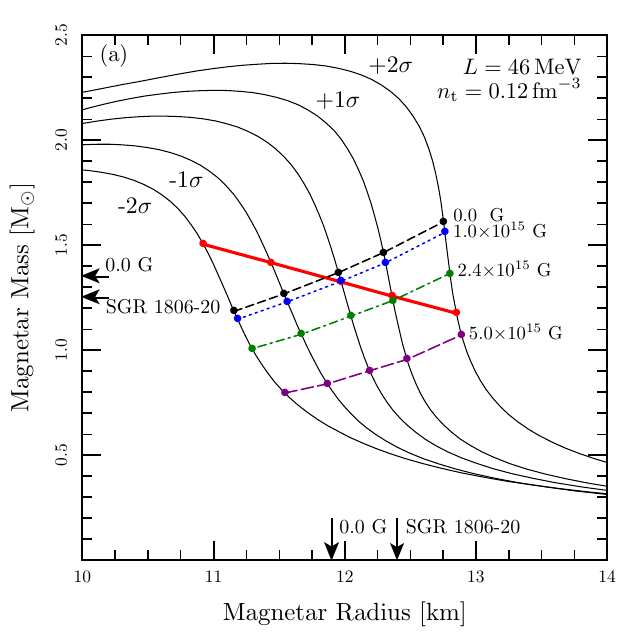}\hspace{0.04\doublewide}
\includegraphics[width=0.48\doublewide]{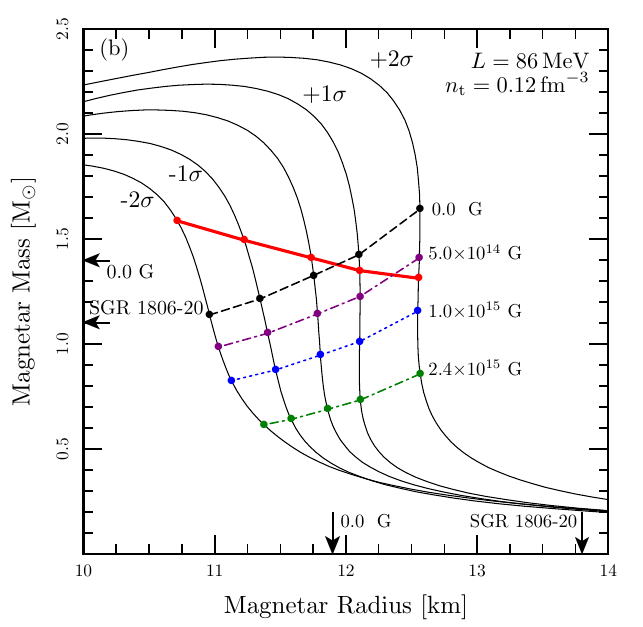}
\caption{(Color online) Magnetar mass as a function of radius for the core EOS
  probability distribution from \citet{steiner12}. Frequencies are
  evaluated using (a) the SLy4 crust EOS ($L = 46 \MeV$) and
  (b) the Rs crust EOS ($L = 86 \MeV$), both
   for $n_{\mathrm{t}} = 0.12\invcfm$. The thick red solid
  line indicates masses and radii for which the fundamental mode has a
  frequency of $29\Hz$ in the case of SLy4 and $18\Hz$ in the case of Rs. 
  The black short-dashed line indicates masses
  and radii for a $626\Hz$ harmonic mode and $B=0\gauss$. Masses and
  radii from $626\Hz$ harmonic modes with magnetized crusts are
  labeled accordingly. Arrows indicate masses and radii that match
  both the fundamental and the harmonic modes for the field-free case
  and the case with the magnetic field of SGR 1806-20
  ($B=2.4\times10^{15}\gauss$). }
\label{f.mass-radius}
\end{figure*}

We examine the sensitivity of fundamental and harmonic modes to
the strength of the magnetic field. Strong magnetic fields melt the
outermost boundary of the crust (i.e., push the melting point of the
one-component plasma to higher pressures). Since $R_{\mathrm{core}}$
remains fixed and $R_{\mathrm{crust}}$ decreases, a strong magnetic
field thins the crust (i.e., decreases $\Delta$). Although a strong
magnetic field can decrease the crust thickness and change the
composition of the outer crust, the overall impact on predicted
fundamental and harmonic mode frequencies is negligible. We find that
predicted fundamental modes from magnetized crusts are nearly
identical to the field-free case, in agreement with the findings
  by \citet{nandi12}. The magnetic field is not a determining factor
because fundamental crust modes are entirely set by our choice of
radius for the magnetar. That is, fundamental modes are entirely set
by the equation of state. However, a magnetized crust can
significantly alter predicted harmonic modes. The $n=1$ modes are
sensitive to the magnetic field, especially in the outer crust where
$\vAlfven > \vshear$ \citep{piro05,nandi12}. 
For example, to obtain a mass-radius solution consistent with the findings of
\citet{steiner12} for the Rs EOS requires $B\lesssim 10^{15}\gauss$.
The magnetized crusts that match
observed QPOs can be seen in Fig.~\ref{f.mass-radius}.

\subsection{The crust-core transition density}

\begin{figure}
\includegraphics[width=\figwidth]{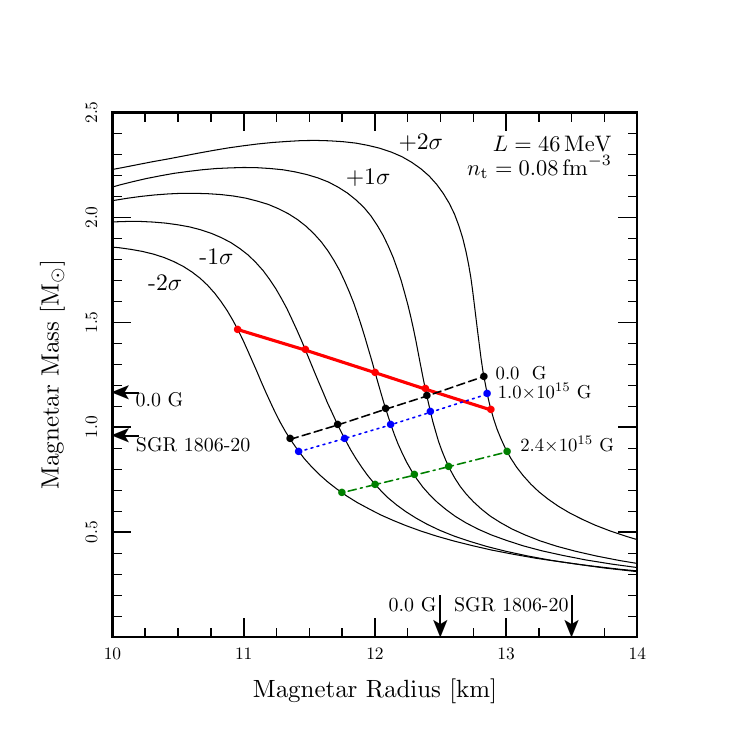}
\caption{(Color online) Same as Fig.~\ref{f.mass-radius}, but for the SLy4 crust
  EOS with $n_{\mathrm{t}} = 0.08\invcfm$. The thick red solid line indicates masses and
  radii determined from a fundamental mode of $29\Hz$. Masses and radii from
  harmonic modes with magnetized crusts are labeled accordingly.
  Arrows indicate masses and radii that match both the fundamental and
  the harmonic modes for the field-free case and the case with the
  magnetic field of SGR 1806-20 ($B=2.4\times10^{15}\gauss$).}
\label{f.mass_radius_Rs}
\end{figure}

We test two extreme values for the crust-core transition density,
$n_\mathrm{t}$, from \citet{oyamatsu07}. The exact value of the
crust-core transition density is unknown, in part because the density
dependence of the nuclear symmetry energy in the inner crust is not
well constrained and also because of the possible existence of nuclear
pasta \citep[see, e.g.,][]{newton12}. Although previous works have
  found correlations between $n_\mathrm{t}$ and $L$ \citep{newton12}, these correlations are still rather
  model-dependent. In Fig.~\ref{f.mass-radius}, we examine various magnetic field strengths to find masses and
radii from intersections of fundamentals and harmonics for each transition
density . For the SLy4 EOS, $n_{\mathrm{t}} = 0.12\invcfm$, and a surface dipole field
matching that of SGR 1806-20 \citep{mcgill}
($B=2.4 \times 10^{15}\gauss$) we find the magnetar to have
$M=1.25\Msun$ and $R=12.4\km$. We must extrapolate outside the
equation of state curves to approximate a mass and radius for the
lower crust-core transition density $n_{\mathrm{t}} = 0.08\invcfm$, 
as can be seen in Fig.~\ref{f.mass_radius_Rs}. This crust-core
transition gives $M=0.96\Msun$ and $R=13.5\km$ for SGR 1806-20. In
either case, if we assume that the magnetic field inside the crust is
larger than the observed surface field, then a smaller mass and larger
radius is implied. If the magnetic field is too large, the implied
radius will be far outside the radii implied by mass and radius
observations from the quiescent LMXBs in M13 and $\omega$~Cen
\citep{steiner-lattimer-brown-10}. 
Using the Rs EOS,
$n_{\mathrm{t}} = 0.12\invcfm$, and a surface dipole field matching that of SGR 1806-20 ($B=2.4
\times 10^{15}\gauss$) we find the magnetar to have $M=1.10\Msun$ and
$R=13.8\km$, which is outside the 2-$\sigma$ mass-radius relation of
\citet{steiner12}, as can be seen in Fig.~\ref{f.mass-radius}.

Figure \ref{f.mass-radius} also demonstrates that only harmonic modes are affected by a change in
the crust-core transition density. A change in the crust-core
transition density will change the crust thickness and harmonic modes
scale with the crust thickness. Fundamental modes remain unchanged,
however, because they scale with the radius of the entire star.

\subsection{The entrainment of the free neutron gas}

\begin{figure}
\includegraphics[width=\figwidth]{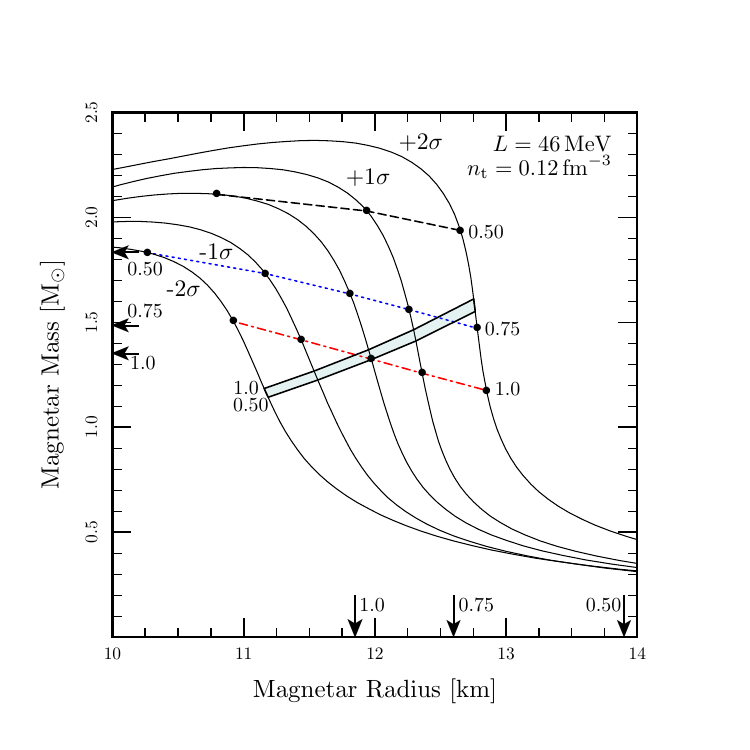}
\caption{(Color online) Magnetar mass as a function of radius for the core EOS probability
  distribution from \citet{steiner12}. Frequencies are evaluated using
  the SLy4 crust EOS with $B = 0\gauss$ and $n_{\mathrm{t}} =
  0.12\invcfm$. The red dot-dashed, blue dotted, and black dashed
  lines indicate masses and radii from fundamental modes of frequency
  $29\Hz$ for different free neutron entrainment fractions
  $\fent$. The shaded band indicates masses and radii from
  $626\Hz$ harmonic modes as $\fent$ is varied from $0.50$
  to $1.0$. Arrows indicate the masses and radii that match both the
  fundamental and the harmonic modes for $\fent=1.0$,
  $0.75$, and $0.50$.}
\label{f.mass-radius-entrainment}
\end{figure}

\begin{figure}
\includegraphics[width=\figwidth]{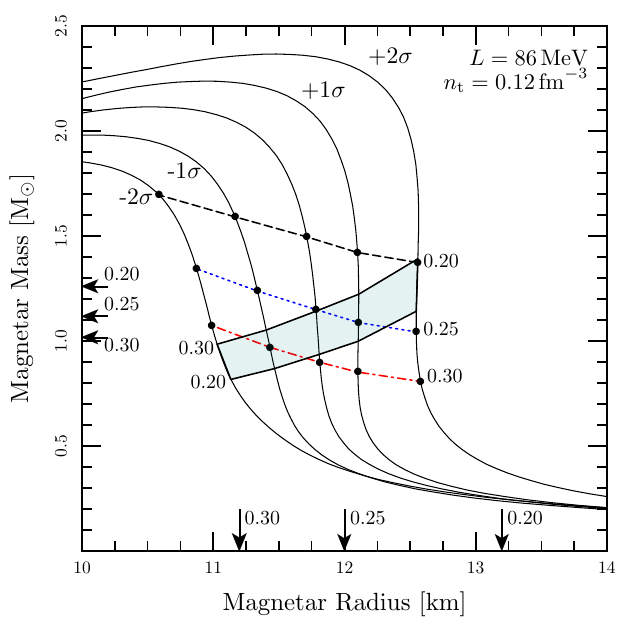}
\caption{(Color online) Same as Fig.~\ref{f.mass-radius-entrainment}, but for the
  Rs crust EOS with $B = 0\gauss$. Here the free neutron entrainment
  fraction $\fent$ is varied from 0.20 to 0.30, with
  $\fent$ labeled next to the corresponding curves. The red
  dot-dashed, blue dotted, and black dashed lines indicate masses and
  radii from the $29\Hz$ fundamental mode. The shaded band indicates
  masses and radii from the $626\Hz$ harmonic mode. Arrows indicate
  the masses and radii for $\fent=0.30$, $0.25$, and $0.20$
  that match both the fundamental and harmonic modes.}
\label{f.mass_radius_Rs_entrainment}
\end{figure}

Entrainment of the free neutron gas in the inner crust alters both
fundamental and harmonic modes, as shown in
Fig.~\ref{f.mass-radius-entrainment}. We define the degree of
entrainment $\fent$ as the fraction of the free neutron gas that moves
with the lattice during a crust oscillation. Although $\fent$ only
slightly alters the harmonic frequency, mainly by changing $\vshear$,
it significantly alters the fundamental mode frequency. This occurs
because the fundamental mode energy is concentrated deeper in the
crust, whereas harmonic modes have their energy distributed more
uniformly over the crust \citep{piro05}. If a lower fraction of the
free neutrons are entrained, then larger masses and radii are implied
for the magnetar. We find that a large degree of entrainment, $\fent >
0.75$, is required for predicted crust modes to match observed QPOs
using crust models built upon the SLy4 interaction~\citep{cha95}.
As shown in Fig.~\ref{f.mass_radius_Rs_entrainment}, for the Rs EOS
the predicted $n=0$, $l=2$ fundamental gives frequencies near the 29~Hz QPO of SGR 1806-20 when there is a low degree of entrainment of the
free neutrons. For example, for $\fent=0.25$ we find the magnetar to
have $M=1.12\Msun$ and $R=12.0\km$.

\section{Discussion}\label{s.discussion}

Magnetar giant flare QPOs provide a unique opportunity to probe the
nuclear physics of the neutron star crust. Fundamental torsional modes are largely
independent of the crust-core transition density and the magnetic
field strength. Harmonic modes are sensitive to the surface gravity,
transition density, entrainment of the free neutrons, and the magnetic
field strength. Comparison of fundamental and harmonic modes gives
solutions for magnetar masses and radii and hence places constraints on $L$. In particular, we find values of $L$ that give results consistent
  with observed oscillations.
For the SLy4 EOS, solutions most consistent with these constraints have large crust-core
transition densities and a large degree of free neutron entrainment; those for the Rs EOS have large crust-core transition densities and a small
degree of free neutron entrainment.

We find, in agreement with
\citet{sotani13}, smaller fundamental mode frequencies for crust equations of state
with larger values of $L$. Both works find that more entrainment
decreases the fundamental frequency (in the notation of
\citet{sotani13}, $N_s/N_d = 1-\fent$). Our work includes nuclear
shell effects in a more consistent fashion, and thus it is more
difficult to vary $L$ continuously as in \citet{sotani13}. Also, we
only employ equations of state that are consistent with recent constraints from
neutron star mass and radius measurements \citep{steiner12} that rule
out larger values of $L$. A complete evaluation of how the entrainment
in the crust might be correlated with $L$ is needed and work in this
direction is in progress.
 
Although fundamental modes are only slightly affected by the
crust-core transition density, a larger transition density increases
the crust thickness, for a fixed mass and radius, and drives the
harmonic frequency lower. To match the observed harmonic with a larger
transition density therefore requires a larger mass for a fixed
radius.

The degree of entrainment of the free neutron gas in the inner crust
alters both fundamental and harmonic modes by changing the shear
velocity \citep{chamel12}. The fundamental mode is most sensitive
  to the entrainment fraction at the highest densities in the crust. A
  recent study of neutron entrainment via Bragg scattering with the
  crystal lattice gives $\fent \approx 0.35\textrm{--}0.90$
  throughout the inner crust \citep{chamel12}. Since the density dependence
  of neutron entrainment is model dependent and has not been studied
  for the equations of state considered here, we assume a
  fixed fraction of neutron entrainment throughout the entire inner crust.
We find that $\fent\gtrsim 0.75$ is required to have modes consistent
with observed QPOs for the SLy4 crust EOS with $L=46\MeV$. For
example, with full entrainment, $\fent=1.0$, we find
$M=1.35\Msun$ and $R=11.9\km$. In contrast, for $\fent=0.5$ the
solution lies outside the 2-$\sigma$ $M$-$R$ relation from
\citet{steiner12}; extrapolating the fundamental and harmonic curves
gives $M=1.83\Msun$ and $R=13.9\km$. The Rs crust EOS with $L=86\MeV$
requires $\fent\lesssim0.30$ to have modes consistent with observed
QPOs. In general, to achieve reasonable values of $R$ with lower
values of $L$ requires a larger $\fent$ and a larger $n_{\mathrm{t}}$.

Although the observed surface dipole field strengths are too weak to
affect the fundamental torsional modes of magnetars, the harmonic
modes are significantly altered by fields $\gtrsim 10^{15}\gauss$. For
a transition density at $0.12\invcfm$ ($0.08\invcfm$) a magnetic field
of $B \gtrsim 4.0 \times 10^{15}\gauss$ ($\gtrsim 2.0\times
10^{15}\gauss$) gives no mass and radius solutions consistent at the
2-$\sigma$ level with the mass and radius constraints from PREs and
LMXBs. For all crust-core transition densities, a magnetized crust
requires a lower mass than the field-free case in order to contain a
mode consistent with the observed 626~Hz QPO. The field-free case
gives a minimum radius for a crust that can reproduce observations of
SGR 1806-20; our model requires $R\geq 11.9\km$ for SGR 1806-20. The
sensitivity of the harmonic modes to the crust magnetic field strength
suggests that the local magnetic field strength cannot greatly exceed
the inferred dipole surface field if the QPOs are identified with
torsional modes for neutron star masses and radii consistent with
those of PREs and LMXBs. We have taken our core EOS models from
\citet{steiner12}, who found that smaller radii were disfavored by
recent neutron star radius measurements. Recent analysis of quiescent
thermal emission from transient neutron stars suggest that the radii
are $< 11.1\km$ (99\%-confidence; \citealt{guillot13}). Matching
torsional modes to observed QPOs might still be possible in this case
if either $L$ or $\fent$ were sufficiently large. This
would also require that magnetars have a rather low mass.

Our analysis assumes that the QPOs are due to torsional modes of the
crust and that the crust is decoupled from the core. That neutrons in
the inner crust would form a superfluid is an idea predating the
discovery of neutron stars \citep{Migdal1959Superfluidity-a}, and
there is both theoretical \citep[see,
  e.g.,][]{Gezerlis2010Low-density-neu} and observational evidence
from cooling transients
\citep{Shternin2007Neutron-star-co,Brown2009Mapping-Crustal} that the
neutrons are below their transition temperature in the inner crust.
The neutron superfluid can plausibly decouple the crust and core by
eliminating viscous drag \citep{ruderman68} and has long been used to
explain the long-relaxation times of pulsar glitches \citep{baym69,
  link93, link12}. In the presence of a magnetic field the crust and
core are not completely decoupled. Indeed, \citet{Gabler11} found that
for $B\gtrsim 5\times 10^{13}\gauss$ torsional crust modes would be
resonantly damped by coupling to the Alfv\'en continuum, with damping
time scales $\sim 0.2\,\mathrm{s}$. Their study did not, however,
include the effects of proton pairing in the core nor did it include a
realistic model of the neutron star crust. Observations of cooling of
the young neutron star in the Cas A supernova remnant
\citep{heinke10,shternin11} suggest that the proton $^{1}S_{0}$
pairing gap is large \citep{page11,yakovlev11}, so that the protons
are in a superconducting state throughout the core. The crust-core
coupling depends on the magnetic field configuration and the magnetic
field strength near the crust-core interface \citep{gabler12}, neither
of which are well understood. Understanding the coupling between shear
modes and magneto-elastic oscillations in the presence of
superfluidity remains challenging. If the threshold field for damping
via coupling to the Alfv\'en continuum were in actuality substantially
larger than $B=5\times 10^{13}\gauss$, then the calculations in this
paper would still apply.

It is also possible that in the presence of superfluidity, axial
perturbations will be pinned to some extent to the core depending on
the strength and configuration of the magnetic field. We note that for
the magnetic fields studied here ($\le 2.4\times 10^{15}\gauss$), the
Alfv\'en velocity is more than an order of magnitude smaller than the
shear velocity at the crust-core interface (see
Fig.~\ref{f.Alfven-shear-speeds}). In this case, magnetic stresses
are likely to be much smaller than elastic stresses, and our findings
are not likely to change.

\begin{acknowledgments} The authors thank Andrew Cumming and
Sanjay Reddy for useful discussions. This work is supported by NASA
ATFP grant NNX08AG76G, U.S. DOE grant DE-FG02-00ER41132, Chandra grant
TM1-12003X, NSF AST grant 11-09176, and by the Joint Institute for Nuclear
Astrophysics at MSU under NSF PHY grant 08-22648.
\end{acknowledgments}

\appendix
\section{The Crust Equation of State}\label{s.crustEOS}

To compute the energy density of matter in the crust $w$, we start
with an expression similar to that used by \citet{baym71}. We take our
crust to be composed of ``drops'' of nuclear matter with volume
fraction $\chi$; within the nucleus the density of neutrons and
protons are $n_{n}$ and $n_{p}$, respectively, and we denote $n_{l} =
n_{n}+n_{p}$ to be the average baryon density inside a nucleus. The
dripped neutrons, with density \ndrip, occupy a fraction $1-\chi$ of
the volume. The density of nucleons per unit volume is thus $n =
\chi(n_{n}+n_{p}) + (1-\chi)\ndrip$, and the density of electrons is
$n_{e}$. As the density approaches nuclear saturation the fraction of
space filled by the neutron gas approaches unity.

The energy density $w$ has contributions from nuclei (including the
Coulomb lattice contribution), dripped neutrons, and electrons:
\begin{eqnarray}
\nonumber
w(Z,A,n) &=& \chi  \left[ n_{n}m_{n} + n_{p}m_{p} + n_{l}\frac{\Ebind(Z,A)}{A}\right]\\
 & &+ (1-{\chi})\epsilon(n_{n} =\ndrip,n_p=0)  +  w_{e}(n_{e}) .
\label{e.w}
\end{eqnarray}
This expression is valid for any baryon density below the transition
density ($\approx10^{14} \, \mathrm{g} \, \mathrm{cm}^{-3}$). Here
$\epsilon(n_{n},n_{p})$ is the energy density, including rest mass, of
homogeneous bulk matter at a given neutron and proton number density.
We compute $\epsilon$ using the bulk matter Hamiltonian in the Skyrme
model \citep{skyrme59} with SLy4 coefficients \citep{cha95}.
 
The energy density of the nucleus is
\begin{eqnarray}
\nonumber
\lefteqn{n_{n}m_{n}+n_{p}m_{p} + n_{l}\frac{\Ebind(Z,A)}{A} =}\\
 & & \epsilon(n_{n},n_{p})  + \frac{n_{l}}{A}\left(\Esurf + \Eshell + \Epair\right) + \wCoul.
\label{e.nuclear-energy-density}
\end{eqnarray}
In this expression, $n_{n}$ and $n_{p}$ are the neutron and proton
densities inside the nucleus. For the nuclear and lattice
contributions to the energy density $\Ebind$, we use a liquid-drop
mass model \citep{baym71,baymp71,ravenhall83,steiner08} that includes
the lattice contribution in the Coulomb term \wCoul, as well as
surface (\Esurf), shell (\Eshell), and pairing (\Epair) corrections to
the homogeneous bulk matter Hamiltonian $\epsilon$. At lower
densities, the energy per particle in the crust is minimized when
$n_{\mathrm{drip}}=0$, and after the neutron-drip point (about $4
\times 10^{11}$ g/cm$^3$), the energy per particle is minimized only
when $n_{\mathrm{drip}}>0$. The baryon number density inside a nucleus
$n_{l}$ is determined from
\begin{equation}\label{e.baryon-density-inside}
n_{l} = n_{0} + n_{2}I^{2},
\end{equation}
where $I = 1 - 2Z/A$ is the isospin asymmetry, $n_{0}$ is the nuclear
saturation density of bulk homogeneous matter, and $n_{2} < 0$ is a
correction due to both the isospin asymmetry, which decreases the
saturation density, and the Coulomb interaction, which increases the
saturation density \citep{steiner08}. The average neutron and proton
densities within the nucleus are then determined from $n_{l}$ and $I$
via
\begin{equation}
n_{n} = \frac{n_{l}}{2}(1+ \eta I),\quad n_{p} = \frac{n_{l}}{2}(1- \eta I),
\end{equation}
where $\eta = \delta/I= 0.92$ is a constant of our model that
determines the thickness of a neutron skin \citep{steiner08}, i.e.,
the difference between neutron and proton radii, and $\delta =
1-2n_{p}/(n_{n}+n_{p})$ is the density asymmetry.

The next three terms in Equation~(\ref{e.nuclear-energy-density}) are the surface, shell, and pairing corrections.
The surface correction is proportional to  the surface tension $\sigma$, the nuclear surface area $A^{2/3}$, and density asymmetry $\delta$,
\begin{equation}
\Esurf = \sigma \left(\frac{36\pi A^{2}}{n_l^2}\right)^{1/3} \left( 1 - \sigma_{\!\delta}{\delta}^{2} \right)
\end{equation}
where $\sigma_{\!\delta} > 0$ is a parameter that represents the surface asymmetry \citep{myers69,steiner05}. 
The shell correction to the binding energy per baryon is \citep{dieperink09} 
\begin{equation}
\Eshell(Z,N) = a_{1}S_{2} + a_{2}S_{2}^{2} + a_{3}S_{3} + a_{np}S_{np},
\end{equation}
where the $a_{i}$ are fitting parameters,
\begin{eqnarray}
S_{2} &=& \frac{n_{v}\bar{n}_{v}}{D_{n}} + \frac{z_{v}\bar{z}_{v}}{D_{z}} , \\
S_{3} &=& \frac{n_{v}\bar{n}_{v}(n_{v}-\bar{n}_{v})}{D_{n}} + \frac{z_{v}\bar{z}_{v}(z_v-\bar{z}_v)}{D_{z}} , \\
S_{np} &=& \frac{n_{v}\bar{n}_v z_{v}\bar{z}_v}{D_{n}D_{z}},
\end{eqnarray}
and
\begin{eqnarray}
\bar{n}_v &\equiv& D_n - n_v ,\\
\bar{z}_v &\equiv& D_z - z_v .
\end{eqnarray}
The parameters $D_n$ and $D_z$ correspond to the degeneracy of the
neutron and proton shells, i.e., the difference between the magic
numbers enclosing the current amount of neutrons or protons. The
quantities $n_v$ and $z_v$ are the number of valence neutrons and
protons, i.e., the difference between the current number of protons or
neutrons and the preceding magic number. The pairing contribution to
the nuclear binding energy is taken from \citet{brehm89} with updated
coefficients,
\begin{equation}
\Epair = \left\{
\begin{array}{cr}
  -a_p A^{-1/3}, & \textrm{even-even} \\ 
 +a_p A^{-1/3}, & \textrm{odd-odd}\\
  0, & \textrm{even-odd} 
\end{array}\right. , 
\end{equation}
where $a_p$ is a constant of our model. The last term in
Equation~(\ref{e.nuclear-energy-density}) is the Coulomb energy
density,
\begin{equation}
\wCoul = \frac{2\pi}{5}n_{p}^{2}e^{2}R_{p}^{2}\left(2-3{\chi}^{1/3}+\chi\right),
\end{equation}
where $e^{2}$ is the Coulomb coupling and $R_p$ is the proton radius
($3Z = 4\pi n_pR_p^{3}$). The respective $\chi$ terms in parentheses
correspond to the Coulomb contribution, the lattice contribution, and
a correction that accounts for the filling fraction $\chi$ of the
nuclei. Table~\ref{t.nuclear-parameters} lists the values of the
coefficients used in this mass model.

\begin{table}
\caption{\label{t.nuclear-parameters} Parameters of the mass model.}
\begin{ruledtabular}
\begin{tabular}{ccc}
parameter & SLy4 & Rs \\
\hline
$n_{0}$ & $0.1740\invcfm$ & $0.1597\invcfm$ \\
$n_{2}$ & $-0.0157\invcfm$ & $0.0244\invcfm$ \\
$\eta$ & 0.9208 & 0.9043 \\
$\sigma_{\!\delta}$ & 1.964 & 1.465 \\
$\sigma$ & $1.164 \MeV$ & $1.041 \MeV$ \\
$a_{1}$ & $-1.217 \MeV$ & $-1.298 \MeV$ \\
$a_{2}$ & $0.0256 \MeV$ & $0.0311 \MeV$ \\
$a_{3}$ & $0.00387 \MeV$ & $0.00349 \MeV$ \\
$a_{np}$ & $0.0357 \MeV$ & $0.0287 \MeV$ \\
$a_{p}$ & $5.277 \MeV$ & $5.265 \MeV$ \\
\end{tabular}
\end{ruledtabular}
\end{table}

The electronic contribution to the energy density is that of an
electron gas embedded in a uniform magnetic field. The electrons
acquire an effective mass $m_f$ in the presence of the magnetic field
\begin{equation}
m_f^2 = m_e^2+ 2m_e^2 \left(x+\frac{1}{2} + \frac{1}{2}\nu\right) B_* ,
\end{equation}
where $m_e$, $x$, and $\nu$ are respectively the electron mass,
principal quantum number, and electron spin along the magnetic field
\citep{Rabi28,Ventura2001}. Here $B_* = \hbar eB/m_{e}^{2}c^{3} =
B/(4.414\times10^{13}\gauss$) is the ratio of the magnetic field to
the critical field, defined as the field at which the cyclotron energy
equals the electron rest-mass. The electron number density and energy
density are found by summing over electron states and spins in the
limit $\mu_e \gg m_f$.

\bibliography{arxiv2}

\end{document}